# Self Consistent Simulation of C-V Characterization and Ballistic Performance of Double Gate SOI Flexible-FET Incorporating QM Effects

Zubair Al Azim[1,2], Nadim Chowdhury[1], Iftikhar Ahmad Niaz[1,2]

[1]Department of Electrical and Electronic Engineering
Bangladesh University of Engineering and Technology,
Dhaka-1000, Bangladesh
Email: zubair_al_azim@ieee.org

Md. Hasibul Alam[1], Imtiaz Ahmed[1,2], Quazi D.M. Khosru[1,2]

[2]Department of Electrical and Electronic Engineering
Green University of Bangladesh,
Dhaka-1207, Bangladesh

*Abstract*— Capacitance-Voltage (C-V) & Ballistic Current-Voltage (I-V) characteristics of Double Gate (DG) Silicon-on-Insulator (SOI) Flexible FETs having sub 35nm dimensions are obtained by self-consistent method using coupled Schrodinger-Poisson solver taking into account the quantum mechanical effects. Although, ATLAS simulations to determine current and other short channel effects in this device have been demonstrated in recent literature, C-V & Ballistic I-V characterizations by using self-consistent method are yet to be reported. C-V characteristic of this device is investigated here with the variation of bottom gate voltage. The depletion to accumulation transition point (i.e. Threshold voltage) of the C-V curve should shift in the positive direction when the bottom gate is negatively biased and our simulation results validate this phenomenon. Ballistic performance of this device has also been studied with the variation of top gate voltage.

## I. INTRODUCTION

Remarkable immunity to short channel effects and excellent device scaling possibilities has led to the development of double gated MOSFETs. Flexible threshold voltage field effect transistor (Flexible-FET) has been reported as one the most scalable [1] independently-double-gated MOSFETs with a unique dynamic threshold voltage control feature using the bottom gate voltage. Flexible-FET is the only widely recognized device that combines a JFET with a MOSFET. It has a damascene metal top gate and an implanted JFET bottom gate that are self aligned in a gate trench [2] (Fig 1). Flexible-FETs have been experimented to show sub-threshold slope near 64mV/decade and $I_{on}/I_{off}$ ratios on the order of $10^5$ [3]. With all its exclusive features, Flexible-FET is being considered as an alternative to FinFET as it has certain fabrication advantages [3].

Flexible-FET was proposed initially at 180 nm channel length [1]. Continuous scaling has led to the evaluation of this device at 32 nm channel length [4]. As device dimensions reach such scales, quantization of carrier energies lead to increased quantum mechanical (QM) effects which manifest themselves in measurable device parameters like threshold voltage, drive current, gate capacitance etc [5]. Hence detailed consideration of QM effects is essential in understanding the electrical and electrostatic behaviors of Flexible-FETs with 32nm channel length. A well established approach of studying the quantum effects resulting from the formation of energy sub-bands in such nano-scale device structures is the self-consistent method where coupled Schrodinger's and Poisson's equations are solved iteratively. Important device characteristics for 32nm Flexible-FETs have been reported using SILVACO simulations [4]. However, self-consistent C-V modeling and Ballistic I-V characterization is yet to be reported for 32nm Flexible-FETs.

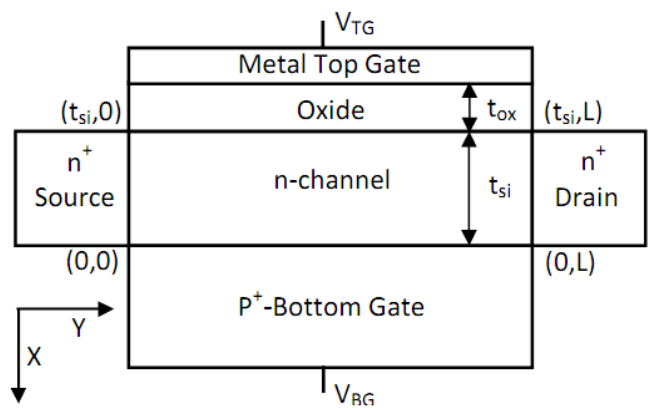

Fig 1: n-channel Flexible-FET top view

In this paper, Capacitance-Voltage characteristic (C-V) of Flexible FETs having sub 35nm dimensions are obtained by self-consistent method using coupled Schrodinger-Poisson solver taking into account the quantum mechanical effects. Ballistic model has been used to obtain I-V characteristic. Variation of device performance with top and bottom gate bias has also been illustrated.

## II. Self Consistent modeling

For determining the gate capacitance of FD and DG SOI Flexible FETs self-consistently, 1D coupled Poisson's and Schrodinger's equations are solved along the direction perpendicular to the direction of transport. The Poisson's equation is given by:

$$\nabla^2 \varphi(x) = -\frac{q}{\epsilon(x)}[p(x) - n(x) + N_D - N_A] \qquad (1)$$

Where $\Phi$, p, n, $N_D$ and $N_A$ are the potential (V), hole concentration ($m^{-3}$), electron concentration ($m^{-3}$), donor atom concentration ($m^{-3}$) and acceptor atom concentration ($m^{-3}$) respectively. q is the electronic charge and $\epsilon(x)$ is the dielectric constant where (x) denotes that it has different values in different layers. Since the silicon film is uniformly doped, $N_D$ and $N_A$ are considered to be constants. The Schrodinger's equation can be written as

$$\left(-\frac{\hbar^2}{2m_{ds}^*}\nabla^2 - q\varphi(x)\right)\Psi_j(x) = E_j\Psi_j(x) \qquad (2)$$

Where $m_{ds}^*$ is the density-of-states (DOS) electron mass and $E_j$ and $\Psi_j$ are the minimum energy and corresponding wave function of the j[th] subband respectively. The DOS mass of electron for bulk silicon is defined as

$$m_{ds}^* = 6^{\frac{2}{3}}(m_t^{*2}m_l^*)^{\frac{1}{3}} \qquad (3)$$

Which is found to be $1.08m_0$ for silicon crystal where $m_0$ is the free electron mass. $m_t^*$ and $m_l^*$ are the transverse and longitudinal electron effective masses in a three dimensional (3D) silicon crystal [6], [7]. Effective mass variation with the silicon crystal orientation is not considered here; however this can be considered by assuming different effective mass values depending on the particular crystal orientation considered. Nonetheless, the qualitative nature of the results and discussions presented here will be equally applicable for either case. Wave function penetration effect is not considered in this work. From the solution, electron concentration is obtained using the wave function and Eigen energy of each subband.

$$n(x) = \sum_j \left[|\Psi_j(x)|^2 \times \int_{E_j}^{\infty} \rho_j(E) f_{FD}(E)\, dE\right] \quad (m^{-3}) \qquad (4)$$

Where $\rho_j(E)$ is the density-of-states (DOS) in the j[th] sub band. The term $f_{FD}(E)$ is obtained from the Fermi-Dirac distribution function, given by the well known formula of

$$f_{FD}(E) = \frac{1}{1 + e^{\left(\frac{E-E_F}{kT}\right)}} \qquad (5)$$

Here $E_F$ is the Fermi energy. The device considered here has quantum confinement in one direction, so the resulting DOS system will be that of a quantum well, which has two degrees of freedom [8]. The DOS function for this case is:

$$\rho_j(E) = \frac{m^*j}{\pi\hbar^2 L_{si}} \qquad (m^{-3}j^{-1}) \qquad (6)$$

Where $L_{si}$ is the thickness of the channel. Having reached the convergence of the self-consistent solver, the total charge concentration per unit area is obtained using

$$Q = q \int n(x) dx \qquad (m^{-2}) \qquad (7)$$

Differentiating Q with respect to gate voltage (Vg) gives the capacitance per unit area.

$$C_g = \frac{dQ}{dV_g} \qquad \left(\frac{F}{m^2}\right) \qquad (8)$$

Current flows in the direction perpendicular to the confining potential. The drain current is sum of the contribution from each subband. The drain current contributed by subband j is [9],

$$\frac{I_{Dj}}{W} = \left[\frac{q}{\hbar^2}\sqrt{\frac{m_{cj}}{2}\left(\frac{k_B T}{\pi}\right)^{3/2}}\right]\left\{\mathcal{F}_{\frac{1}{2}}\left[\frac{E_F - E_j}{k_B T}\right] - \mathcal{F}_{\frac{1}{2}}\left[\frac{E_F - E_j - qV_D}{k_B T}\right]\right\} \qquad (9)$$

Where $m_{cj}$ is the conductivity effective mass of subband, j, $E_j$ is the subband energy, $E_F$ is the Fermi level of the source electrode and $\mathcal{F}_{\frac{1}{2}}(x)$ is the Fermi-Dirac integral of order one-half as defined by [10],

$$\mathcal{F}_{\frac{1}{2}}(\eta) = \int_0^{\infty} \frac{\sqrt{y}}{1 + e^{y-\eta}}\, dy \qquad (10)$$

## III. Results and discussion

In this study, silicon channel thickness is considered to be 30 nm. For such small scaled devices, the potential wells become sufficiently narrow to give rise to splitting of the energy levels into sub-bands, which in effect quantizes the carriers at the voltages of interest [11]. Here, carrier confinement is one-directional in the channel region with oxide gate on one side and the p-n junction barrier on the other (Fig 2). The first three quantized states are illustrated in Fig 3.

The capacitance-voltage characteristic is observed for zero and negative bottom gate bias (Fig 4). Variation of total charge is considered for zero and negative bottom gate bias (Fig 5). On the application of a negative bottom gate bias, the p-n junction becomes reverse biased and the total charge in the channel is decreased. The decrease in charge leads to reduced capacitance for negative bottom gate bias.

The ratio of peak value of the electron concentration profile (npeak) and average electron concentration (navg) i.e. npeak/navg is plotted against the top gate voltage (Fig 6). At threshold voltage, the curve shows a distinct transition point which signifies the start of channel charge buildup at a high rate.

The ballistic current is observed with zero bottom gate bias at various top gate bias (Fig 7). With increasing top gate bias there is expected increase in the drain current. Effect of varying oxide thickness on ballistic current is considered in Fig 8. For a smaller oxide thickness the electric field in the channel is increased which results in the increment of ballistic current.

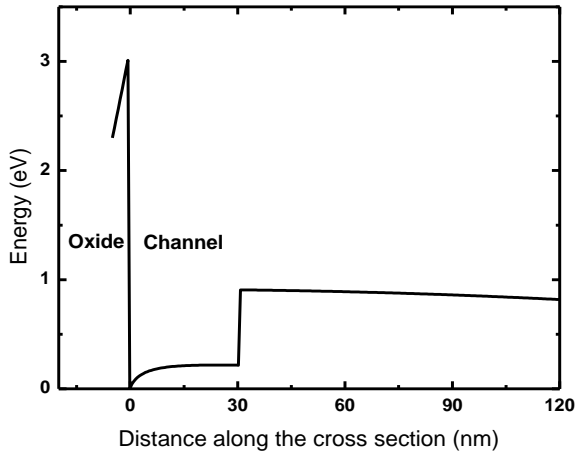

Fig 2: n-channel Flexible-FET carrier cofinement in the channel region

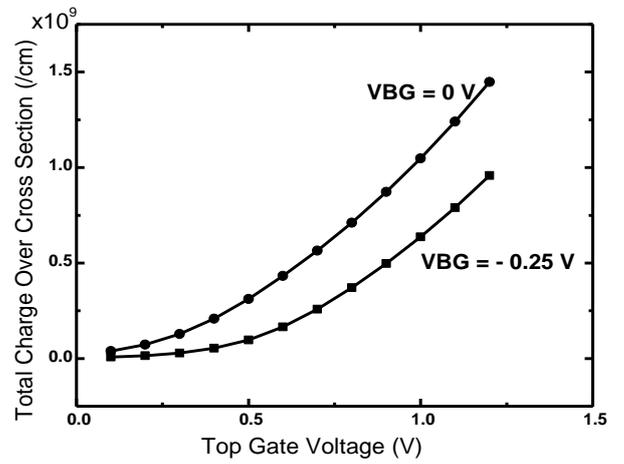

Fig 5: Total charge over the cross section for an n-channel Flexible-FET at various top gate bias

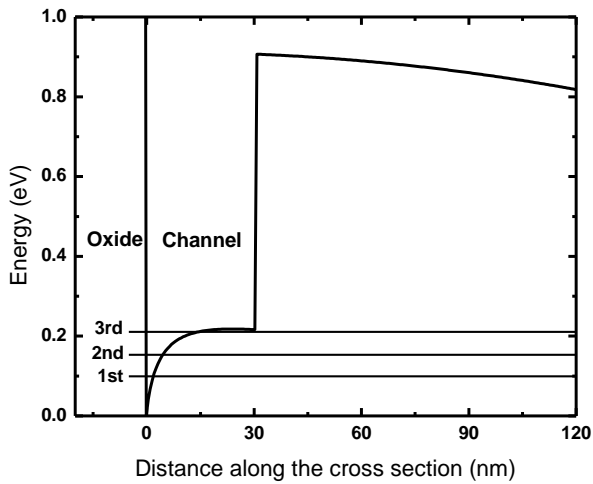

Fig 3: 1st, 2nd and 3rd quantized Energy levels of n-channel Flexible-FET

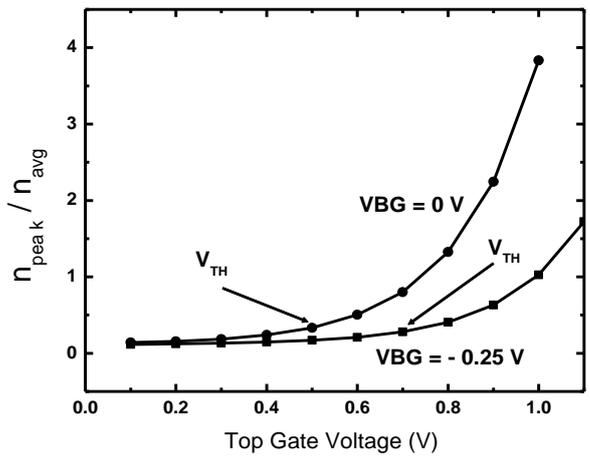

Fig 6: $n_{peak}/n_{avg}$ vs top gate voltage curve for an n-channel Flexible-FET at different bottom gate bias illustrating the shift in threshold voltage

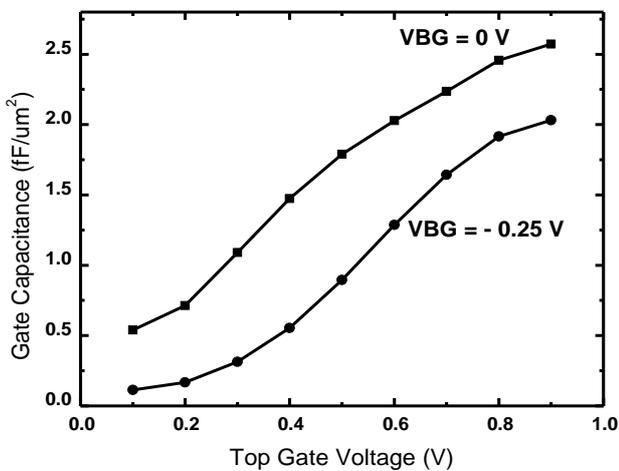

Fig 4: Capacitance-Voltage charectaristic for an n-channel Flexible-FET at different bottom gate bias

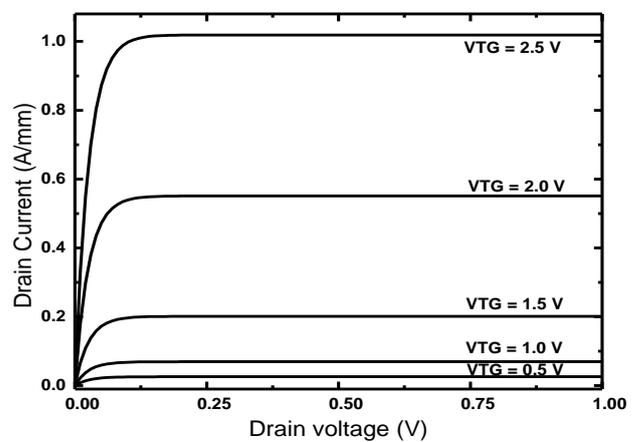

Fig 7: I-V charectaristic for an n-channel Flexible-FET at various top gate bias

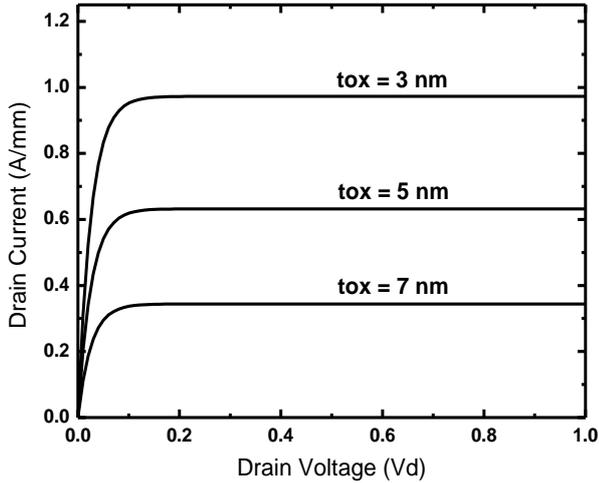

Fig 8: I-V charectaristic for an n-channel Flexible-FET at different oxide thickness levels

## IV. Conclusion

Self consistent simulation method has been applied for C-V characterization of DG SOI Flexible FET for the first time in literature taking into account quantum mechanical effects. Ballistic performance of this device has also been investigated. Our simulation results are in good agreement with the ATLAS simulation results both qualitatively and quantitatively.

## Acknowledgment

We would like to thank Md. Zunaid Baten and Raisul Islam for their support regarding simulations.